\documentclass[aps,floats,amssymb,prb]{revtex4}
\usepackage[unicode,linktocpage=true]{hyperref}
\hypersetup{colorlinks=true,linkcolor=blue,urlcolor =blue,citecolor=blue,anchorcolor=blue} 
\def\dj{d\kern-0.4em\char"16\kern-0.1em}
\def\Dj{\mbox{\raise0.4ex\hbox{-}\kern-0.4em D}}
\usepackage{graphicx}

\begin{document}

\title{Model interactions for Pfaffian paired states based on Chern-Simons field theory description}

\author{Stevan \Dj ur\dj evi\'c$^{1,2}$ and  Milica V. Milovanovi\'c$^{3}$} 
\affiliation{$^1$  University of Montenegro, Faculty of Natural Sciences and Mathematics, D\v zord\v za Va\v singtona bb, 81000 Podgorica, Montenegro}
\affiliation{$^2$ Faculty of Physics, University of Belgrade, Studentski trg 12, 11158 Belgrade, Serbia}
\affiliation{$^{3}$ Scientific Computing Laboratory, Center for the Study of Complex Systems, Institute of Physics Belgrade, University of Belgrade,
Pregrevica 118, 11080 Belgrade, Serbia}

\begin{abstract}
On the basis of Chern-Simons field-theoretical description we propose a simple method for derivation of model interactions for Pfaffian paired states. We verify the method in the case of Pfaffian (i.e. Moore - Read) state, and derive a general form of the model interaction in the case of PH Pfaffian. More than one Landau level is needed to establish the correlations of the PH Pfaffian, and we present the values of relevant three-body pseudo-potentials for two Landau levels.
\end{abstract}

\maketitle

\section{Introduction}

The discovery of the fractional quantum Hall effect (FQHE) at even-denominator, 5/2, filling factor [\onlinecite{will}] initiated an intensive search for viable paired states for the explanation of the effect. A new paradigm of Pfaffian paired states (with $p$-wave Cooper pairs) was introduced  [\onlinecite{mr}] that was the most important building block for the proposals  of topological superconductivity and topological quantum computing. Still, after so many years of intensive experimental and theoretical research, we are not sure which paired state can be associated with the observed even-denominator FQHE.  A paired state, more precisely the theoretical concept of so-called PH Pfaffian, seems according to the recent experiment [\onlinecite{exp}] and other experimental data and theory [\onlinecite{mafe}], and theoretical proposals  [\onlinecite{ph1, ph2, ph3,mafe}], very relevant for the solution of the puzzle at 5/2, though another proposal  [\onlinecite{ss}] was also made that the measured thermal conductance in the experiment of Banerjee et al. ([\onlinecite{exp}]) is a result of an unsufficient equilibration  of the edge modes of  anti-Pfaffian  (a state related to the Pfaffian state).

The PH Pfaffian topological phase may be a result of a disorder dominated physics [\onlinecite{zf, ph1, ph2, ph3}], but might be also a result of Landau level (LL)  mixing in a system that may be considered uniform [\onlinecite{zf, avm}]. It is desirable to understand whether the PH Pfaffian (state) can be supported in a uniform system, as it was done and demonstrated in numerical experiments [\onlinecite{rh, rh1, rh2, rh3, dh, dh1, dh2, dh3, pp, dh4, dh5, rez, phpet}] for Pfaffian and its particle-hole (PH) conjugated partner, anti-Pfaffian [\onlinecite{ap1, ap22}],  taking also into account  their model interactions [\onlinecite{gww, rh}] . The search was proven difficult, because it was found in numerical experiments [\onlinecite{mish}] that the projection of what is believed to be, an appropriate model wave-function of PH Pfaffian, to a fixed LL, represents a gapless state (not a
gapped one necessary for FQHE). Therefore, for numerical experiments and in general, it is very desirable to find a model interaction for the uniform PH Pfaffian state, and in this work we will describe its general form.

The proposal of the PH Pfaffian state is connected with the advance [\onlinecite{son}] in the effective (field-theoretical) description of the Fermi-liquid-like state of composite fermions (underlying quasiparticles) at compressible, even-denominator fractions. Namely, instead of classical composite fermions, we may use Dirac composite fermions to describe these fractions. The PH Pfaffian constitutes a $p$-wave pairing in the opposite sense of the direction dictated by external magnetic field that is materialized in the chiral motion of charge on the edge. The PH Pfaffian also makes the most natural (underlying $s$-wave) pairing of Dirac composite fermions. Thus the solution of the enigma of PH Pfaffian may help us to understand better pairing and superconductivity in Dirac systems raging from the FQHE ones to graphene and topological insulators.

The paper is organized as follows: In the following section, Section II, the Chern-Simons (CS) description is reviewed emphasizing its part responsible for the $p$-wave pairing. In Section III, the method is introduced, based on the CS description and the special pairing part, for establishing model interactions for Pfaffian and PH Pfaffian. The model interaction(s) depend on the sign and strength of a pairing parameter, and, in the last section, Section IV, we analyze and discuss an effective phase diagram and emergent phases as the value of the pairing parameter is varied. The conclusions are in the same section.

\section{Chern-Simons description and Pfaffian paired states}

In the following we will briefly review reasons for Dirac composite fermion theory and the role of  mass in this theory. As argued in [\onlinecite{son}],
the physics of the PH symmetric, half-filled, fixed LL of classical electrons may be connected with the physics of half-filled, $n = 0$ LL of Dirac electrons (in the presence of magnetic field). The PH transformation on real electrons may be considered as a CT (charge conjugation and time reversal) transformation  in this Dirac system. The symmetry under this (CT) transformation is also realized in its dual theory (with fermions that do not couple directly to the external fields)  i.e. Dirac composite fermion theory. The presence of a mass term breaks this symmetry (the mass term transforms into minus itself). Therefore, the mass term in the Dirac composite fermion theory (i.e. its extension with a mass), breaks the PH symmetry of the beginning (classical) electrons, and may mimic LL mixing of  real systems.

The Dirac composite fermion theory [\onlinecite{son}]  can provide a framework for an analysis of Pfaffian paired states as shown in  Ref.  [\onlinecite{avm}].
The  theory  predicts, in the absence of the Dirac mass, equal weight superposition of Pfaffian and anti-Pfaffian, and for small mass either Pfaffian or anti-Pfaffian depending on the sign of mass.  On the basis of the same theory (A) the {\em criticality} of the PH Pfaffian (reversed chirality p-wave) for the {\em zero} Dirac mass case (i.e. in the presence of the PH symmetry) was predicted  [\onlinecite{mvm}, \onlinecite{avm}], that is numerically supported  (no clear gapped state for the two-body interaction in a fixed LL  for PH
symmetric shift on the sphere [\onlinecite{nr}] and high overlap of the projected to the lowest LL usual PH Pfaffian wave function with composite fermion liquid wave function [\onlinecite{mish,bar}]), and (B) the {\em stabilization} was predicted of the usual PH Pfaffian,  i.e. with complex conjugated Pfaffian part of the Pfaffian state, for {\em non-zero} mass. The non-zero mass means the absence of PH symmetry and can mimic LL mixing.  The PH Pfaffian state that follows from the Dirac composite fermion description with a mass [\onlinecite{avm}] is
\begin{equation}
\Psi_{ZF} =   Pf \{ \frac{1}{(z_{i}^* - z_{j}^*)} \}\prod (z_{k} - z_{l})^2 .
 \label{zf}
\end{equation}
 Here
\begin{eqnarray}
Pf\{ \frac{1}{(z_i^* - z_j^*)} \} \sim \nonumber \\
 \sum_P sgn \; P \prod_{i=1}^{N/2} \frac{1}{(z_{P(2 i - 1)}^* - z_{P(2 i)}^*)},  \label{cPf}
\end{eqnarray}
where the sum is over all permutations $(P)$ of $N$ integers.

The field theories (CS or Dirac composite fermion) via gauge fields encode basic interactions and influence among electrons. By a gauge field we describe and summarize the combined effects of Coulomb interaction and constrained dynamics resulting from the fact that electrons mostly live in a fixed LL. The CS  description does not include the projection to a fixed LL, and we will use this description to estimate qualitatively  the influence of other LLs (beyond the first order (in the LL mixing parameter) in the perturbation theory),
when a  system supports a paired state.
(The Ref. [\onlinecite{sm}] gives the first order corrections in the perturbation theory.)

To set a stage and notation we will first review the CS description at filling 1/2 i.e. the HLR description [\onlinecite{hlr}], with non-relativistic composite fermions. This description may be considered as a large $m$ (mass ) limit [\onlinecite{psv,wc}] of the Dirac composite fermion description (which is manifestly symmetric under particle-hole exchange). Thus the non-relativistic description breaks PH symmetry and thus includes  the LL mixing  which promotes the PH Pfaffian (according to Refs. [\onlinecite{zf,avm}]) and makes a natural framework for the investigation of the PH Pfaffian.

We start with the one-particle Hamiltonian,
\begin{equation}
H = \frac{\Psi^\dagger (\bold{p} - \bold{A})^2 \Psi }{2 m} ,
\label{cs1}
\end{equation}
 with  $ c=1, e=1$, and  $\hbar = 1,$ and where $ A_\alpha = -(1/2) B \epsilon_{\alpha \beta} x_{\beta}$ i.e.  $ A_{x} = - (B/2) y $ and $A_{y} = (B/2)  x $, $ \bold{B} = B \mathbf{z} $. We also take $l_B = \sqrt{\frac{\hbar c}{e B}}  = 1$.  The one-particle eigenstates are $ \Psi_m \sim z^m \exp\{-(1/4) |z|^2\} ; m = 0, 1, 2, \ldots $ i.e.  holomorphic  functions (functions only of $z$)  if we do not consider the exponential, $ \exp\{-(1/4) |z|^2\}$ . The CS transformation introduces gauge field $\mathbf{a}$,
\begin{equation}
H = \frac{\Psi^\dagger_{cf} (\bold{p} - \bold{A} - \bold{a})^2 \Psi_{cf}}{2 m} ,
\label{cs2}
\end{equation}
where $ \mathbf{\nabla}  \times \mathbf{a} = - 2 \; \Psi^\dagger  \Psi = - 2 \; \rho(\mathbf{r}) $. In the Coulomb gauge, $ \mathbf{\nabla} \cdot \mathbf{a} = 0$,
\begin{equation}
a_{x} (\mathbf{r}) = 2 \int d \mathbf{r}^{'}  \frac{y - y^{'} }{|\mathbf{r} - \mathbf{r}^{'} |^2} \rho(\mathbf{r}^{'} ),
\label{ax}
\end{equation}
and
\begin{equation}
a_{y} (\mathbf{r}) = - 2 \int d \mathbf{r}^{'}  \frac{x - x^{'} }{|\mathbf{r} - \mathbf{r}^{'} |^2} \rho(\mathbf{r}^{'} ).
\label{ay}
\end{equation}
We would like  to understand the pairing effect of the so-called statistical interaction term,
\begin{equation}
V_{st} = - \mathbf{a} \cdot \frac{\Psi^\dagger_{cf}( \mathbf{p} \Psi_{cf}) -( \mathbf{p} \Psi^\dagger_{cf})  \Psi_{cf} }{2 m} = - \mathbf{a}  \cdot \mathbf{j}_{cf}.
\label{Vst}
\end{equation}
After simple steps and substitutions, which we described in Appendix A, we can arrive at the following expression for the Cooper channel in the inverse space
[\onlinecite{gww}], where operators $a_{\mathbf{p}} $ are associated with the inverse space:
\begin{equation}
V_{st}^{C}  = i \frac{4 \pi }{m} \frac{1}{V} \sum_{\mathbf{q}, \mathbf{p}}  \frac{ (\mathbf{p} \times \mathbf{q} )}{|\mathbf{p} - \mathbf{q}|^2}    a^\dagger_{\mathbf{q}} 
a_{\mathbf{p}}   a^\dagger_{-\mathbf{q}} 
a_{-\mathbf{p}} .  \label{pairint}
\end{equation}
Using complex notation for vectors $\mathbf{p}$ and $\mathbf{q}$ we can rewrite the Cooper channel as
\begin{equation}
V_{st}^{C}  =   \frac{2 \pi }{m} \frac{1}{V} \sum_{\mathbf{q}, \mathbf{p}}  \frac{ (p^* \;  q - p \;  q^*)}{|\mathbf{p} - \mathbf{q}|^2}    a^\dagger_{\mathbf{q}} 
a_{\mathbf{p}}   a^\dagger_{-\mathbf{q}} 
a_{-\mathbf{p}} .
\end{equation}
The second term in $ (p^* \;  q - p \;  q^*)$  with negative sign has a potential to develop pairing instability. We can rewrite that term as
\begin{equation}
\delta V_{st}^{C}  = -  \sum_{\mathbf{q}, \mathbf{p}}  \exp\{i (\theta_p  - \theta_q )\} F    a^\dagger_{\mathbf{q}} 
a_{\mathbf{p}}   a^\dagger_{-\mathbf{q}} 
a_{-\mathbf{p}} ,
\end{equation}
where $F$ is a positive function. Doing the mean field analysis as in Ref. [\onlinecite{gww}] with the effective interaction,
\begin{equation}
\delta_{mf} V_{st}^{C}  = -  \sum_{\mathbf{q}, \mathbf{p}}  \exp\{i (\theta_p  - \theta_q )\} F  \{  < a^\dagger_{\mathbf{q}} 
   a^\dagger_{-\mathbf{q}} >
a_{-\mathbf{p}}  a_{\mathbf{p}} +   a^\dagger_{\mathbf{q}} 
   a^\dagger_{-\mathbf{q}} 
< a_{-\mathbf{p}} a_{\mathbf{p}} > \},
\end{equation}
and using the form of the BCS reduced interaction as in the Ref. [\onlinecite{rg}],
\begin{equation}
\delta_{mf} V_{st}^{C} =   \sum_{ \mathbf{p}} \{ \Delta_{\mathbf{p}}^*  a_{-\mathbf{p}}  a_{\mathbf{p}} +   \Delta_{\mathbf{p}}  a_{\mathbf{p}}^{\dagger}  a_{-\mathbf{p}}^{\dagger} \} ,
\end{equation}
we find for the wave function of the Cooper pair the following behavior,
\begin{equation}
g(\mathbf{p}) \sim \frac{1}{\Delta_{\mathbf{p}}^*} \sim \frac{1}{p} ,
\end{equation}
and in the real space,
\begin{equation}
g(\mathbf{r})  \sim \frac{1}{z} .
\end{equation}
This leads to holomorphic Pfaffian state, because also the basis of single-particle states is holomorphic (up to the exponential factor, functions only of $z$, not $z^*$) .  If there were minus sign
in front of the pairing interaction in (\ref{pairint}) or plus sign instead of minus sign in  (\ref{Vst}),  we would have anti-holomorphic pairing part and this would lead to the PH Pfaffian state (\ref{zf}).

\section{Model interactions for Pfaffian and PH Pfaffian}

Similarly to what was done in Ref. [\onlinecite{gww}] we start with a BCS-like description of the effective pairing interaction for ordinary, non-Dirac composite fermions. Thus we start with
classical composite fermions assuming that the effective mass is large (considerable) and includes the particle-hole symmetry breaking necessary for stabilization / development of Pfaffian/ PH Pfaffian. The effective description we assume is a possible reduction of the CS description with higher order terms, when a $p$-wave state (topological superconductivity) of composite fermions  is established. Thus our beginning Hamiltonian is
\begin{equation}
H_{BCS}^{ef} = \frac{1}{2 m} \Psi_{cf}^\dagger (\mathbf{p})^2 \Psi_{cf} + \lambda \delta \mathbf{a} \mathbf{j}_{cf} ,
\label{bHam}
\end{equation}
where $\delta \mathbf{a} = \mathbf{A} + \mathbf{a}$, and $\mathbf{a}$ is described in Eqs. (\ref{ax}) and (\ref{ay}). Note that we included the regularized form, discussed in the Appendix A, of the statistical interaction i.e. the form with  $\delta \mathbf{a} $ instead of the one with $ \mathbf{a} $
in the previous Section II. This amounts to the substruction of the zero-point energy i.e. (orbital) cyclotron energy due to the motion in a constant magnetic field, which we do not expect to be present in the description of pairing.
The $\lambda $ is an effective coupling which can be negative in the case of Pfaffian and positive for the PH Pfaffian (compare with the discussion in Section II). (The effective BCS interaction
in the case of PH Pfaffian, with $  \lambda > 0 $, is the one that was derived in the scope of the Dirac composite fermion theory  ([\onlinecite{son,psv,wc}]), in the presence of large Dirac mass (i.e. LL mixing, of both signs), in Ref.  [\onlinecite{avm}].)
We rewrite $ H_{BCS}^{ef} $ as
\begin{equation}
H_{BCS}^{ef} = \frac{1}{2 m} \Psi_{cf}^\dagger (\mathbf{p} - \delta \mathbf{a})^2 \Psi_{cf} - \frac{1}{2 m} (\delta \mathbf{a} )^2 \Psi_{cf}^{\dagger} \Psi_{cf}  + (1 + \lambda) \delta \mathbf{a} \mathbf{j}_{cf} .
\label{reHam}
\end{equation}
Now we apply the CS transformation in reverse, from composite fermions to electron representation, to arrive at 
\begin{equation}
H_{BCS}^{el} = \frac{1}{2 m} \Psi^\dagger (\mathbf{p} -  \mathbf{A})^2 \Psi - \frac{1}{2 m} (\delta \mathbf{a} )^2 \Psi^{\dagger} \Psi  + (1 + \lambda) \delta \mathbf{a} \mathbf{J}_{el}  + (1 + \lambda ) \frac{1}{m}  (\delta \mathbf{a} )^2 \Psi^{\dagger} \Psi   .
\label{Ham}
\end{equation}
To get $ H_{BCS}^{el}$ from  $ H_{BCS}^{ef}$  we used the fact that the CS is a unitary, phase transformation [\onlinecite{zhang}] on fermion fields, which transforms the first (kinetic) term in (\ref{reHam})  back to the first term in (\ref{Ham})  (compare (\ref{cs1})  and (\ref{cs2})  in the previous section),
and does not change the form of the second term in (\ref{reHam}) . The third term in (\ref{reHam})   has the composite fermion current, $\mathbf{j}_{cf}$, described in (\ref{Vst}), which transforms as 
\begin{equation}
\mathbf{j}_{cf} \rightarrow \mathbf{j}_{el} + \frac{ \mathbf{a}}{m} \Psi^\dagger \Psi .
\label{cfc}
\end{equation}
Applying this transformation we get the two last terms in (\ref{Ham}) where
\begin{equation}
\mathbf{J}_{el} =  \frac{ - i}{2 m} { \Psi^\dagger (\mathbf{\nabla}  + i \mathbf{A} ) \Psi  - [ (\mathbf{\nabla}  + i \mathbf{A} ) \Psi ]^\dagger  \Psi },
\label{gc}
\end{equation}
is the gauge invariant electron current.
The Hamiltonians $ H_{BCS}^{el}(\lambda) $ describe the effective interactions of electrons that lead to paired states.

The CS description is, in an essence, the Laughlin ansatz (or organization of the solution) translated into the language of field theory, and thus referring mostly to the lowest LL physics. Our goal is a representation of the effective Hamiltonian for paired states in a fixed LL, and, in the following we will use the  lowest LL as a stage for that goal. {\it Thus we will model interactions also for the paired states in the second LL, like Pfaffian, by effective parameters that we will find in the lowest LL.}

To begin modeling in a fixed LL we neglect  (i.e.  consider as a constant) the first term - kinetic term in Eq. (\ref{Ham}). The remaining terms i.e. an effective interaction that we will project to the lowest LL is
\begin{equation}
V_{BCS}^{el}(\lambda) =  (1 + \lambda) \delta \mathbf{a} \mathbf{J}_{el}  + (1/2 + \lambda ) \frac{1}{m}  (\delta \mathbf{a} )^2 \Psi^{\dagger} \Psi  .
\label{Int}
\end{equation}
In the following we will consider the resulting two-body interactions from (\ref{Int}) less important for the physics of paired states, and thus concentrate on the resulting three-body interaction. The contributions from the two-body part have to be carefully calculated, and, 
in Appendix B, the contribution from the first term in (\ref{Int}),  $ \sim  \delta \mathbf{a} \mathbf{j}_{el} $, where $\mathbf{j}_{el} =  \frac{ - i}{2 m} \{ \Psi^\dagger \mathbf{\nabla}  \Psi  -  \mathbf{\nabla}   \Psi^\dagger  \Psi\}$, can be found.



The three-body interaction from the complete effective interaction,  in (\ref{Int}),  is 
\begin{equation}
V_{BCS}^{3}(\lambda) =   (1/2 + \lambda ) \frac{1}{m} : ( \mathbf{a} )^2 \Psi^{\dagger} \Psi :  .
\label{Int3}
\end{equation}
Plugging in the expressions for $\mathbf{a}$ in (\ref{ax}) and (\ref{ay}), we get
\begin{equation}
V_{BCS}^{3}(\lambda) =   (1/2 + \lambda ) \frac{4}{m} \int d \mathbf{r}_1 \int d \mathbf{r}_2     \frac{(\mathbf{r}_3 - \mathbf{r}_1 )  (\mathbf{r}_3 - \mathbf{r}_2 )}{|\mathbf{r}_3 - \mathbf{r}_1 |^{2} |\mathbf{r}_3 - \mathbf{r}_2 |^{2}}: \rho(\mathbf{r}_1 )   \rho(\mathbf{r}_2 )   \rho(\mathbf{r}_3 ) :  ,
\label{Int3explicit}
\end{equation}
i.e. the three-body interaction in coordinate representation is
\begin{equation}
V (\mathbf{r}_1 , \mathbf{r}_2 , \mathbf{r}_3 ) =   (1/2 + \lambda ) \frac{4}{m}    \frac{(\mathbf{r}_3 - \mathbf{r}_1 )  (\mathbf{r}_3 - \mathbf{r}_2 )}{|\mathbf{r}_3 - \mathbf{r}_1 |^{2} |\mathbf{r}_3 - \mathbf{r}_2 |^{2}}. \label{3V}
\end{equation}
The fully antisymmetric wave functions for three particles are given in [\onlinecite{laugh}] and they are:
\begin{equation}
\Psi_{k,l} (\mathbf{r}_1 , \mathbf{r}_2 , \mathbf{r}_3 ) =    \frac{1}{Z_{kl}}   (z_a^2 + z_b^2 )^k  \;  [\frac{(z_a + i z_b )^{3 l} - (z_a - i z_b )^{3 l}}{2 i}] \; \exp\{ - \frac{1}{4} (|z_a |^2  +
|z_b |^2 + |z_c |^2) \} ,  \label{3wf}
\end{equation}
where integers $ k \geq 0 ; \;  l \geq 1 $, and the total angular momentum of the state is $ M = (2 k + 3 l ) $. The normalization factor  $ Z_{k l} = 2^{3 l + 2 k + 1} [ \pi^3 (3 l + k ) ! k ! ]^{1/2} $, and the complex coordinates are
\begin{equation}
z_a = \sqrt{\frac{2}{3}} ( \frac{z_1 + z_2}{2} - z_3 ), \;\;\; z_b = \frac{z_1 - z_2 }{\sqrt{2}}, \;\;\; {\rm and } \; \;  z_c = \frac{z_1 + z_2 + z_3}{\sqrt{3}} .
\end{equation}
Thus $ \sum_{i=1}^{3} |z_i |^2 = |z_a |^2 + |z_b |^2 + |z_c |^2 $.

To describe relevant three-body pseudo-potentials (PPs) we introduce matrix elements of a rescaled three-body interaction: $ \frac{V (\mathbf{r}_1 , \mathbf{r}_2 , \mathbf{r}_3 ) }{\Lambda} $ where  $ \Lambda \equiv (1/2 + \lambda ) \cdot 4/m $.
The diagonal  matrix elements are defined as
\begin{equation}
\Delta_{M = 2 k + 3 l}  = \int d \mathbf{r}_1 \int d \mathbf{r}_2 \int d \mathbf{r}_3 \frac{V (\mathbf{r}_1 , \mathbf{r}_2 , \mathbf{r}_3 ) }{\Lambda} | \Psi_{k,l} (\mathbf{r}_1 , \mathbf{r}_2 , \mathbf{r}_3 ) |^2  .
\end{equation}
The resulting three-body PPs for a fixed $\lambda $ are 
\begin{equation}
V_M (\lambda) = \Lambda   \Delta_M =  (1/2 + \lambda ) \cdot 4/m  \cdot \Delta_M
\end{equation}

The matrix elements relevant for the interaction in the lowest LL are listed in the Table I.
\begin{table}[h!]
\begin{center} 
\caption{Matrix elements in the lowest Landau level.}
\begin{tabular}{|c||c|c|c|c|c|c|c|}
\hline
M & 3 & 5 & 6 & 7 & 8 & 9 & 10 \\
\hline\hline
$\Delta_M$ & $1/24$ & $1/48$ & $7/240$ & $1/80$ & $2/105$ & 
\begin{tabular}{l|l}
$221/10080$ & $1/(240\sqrt{21})$\\
\hline
$1/(240\sqrt{21})$ & $1/120$
\end{tabular} & $3/224$ \\
\hline
$\frac{\Delta_M}{\Delta_{M=3}}$ & $1$ & $0.5$ & $0.7$ & $0.3$ & $\sim 0.475$ & 
\begin{tabular}{l|l}
$\sim 0.526$ & $\sim 0.022$\\
\hline
$\sim 0.022$ & $0.2$
\end{tabular} & $\sim 0.321$ \\
\hline
\end{tabular}
\end{center}
\end{table}
In the table shown are the rescaled values of the three-body PPs of the interaction defined in (\ref{3V})  i.e. $\Delta_M = \frac{V_M (\lambda) }{\Lambda}$, as functions of the total angular momentum, $M$. There are two (orthogonal) wave functions for three fermions at $M = 9$, and thus the corresponding  cases with $ l = 3, k = 0 $, and $ l = 1, k = 3 $,  in the two columns, respectively for $M = 9$, and  a $2 \times 2$ matrix in that subspace.

Remarkably, ratios among the first three values of three-body PPs, $\frac{V_{M = 5}}{V_{M = 3}} =  \frac{\Delta_{M = 5}}{\Delta_{M = 3}} = 0.5 $ and $ \frac{V_{M = 6}}{V_{M = 3}}  = \frac{\Delta_{M = 6}}{\Delta_{M = 3}} = 0.7 $, are quite close to the ones obtained by the first order perturbation theory in the second LL [\onlinecite{sm}], $ \sim 0.4 $ and $\sim 0.7$, respectively. We should notice that in this case the corresponding unit can be expressed as 
$ \frac{e^2}{l_B} \frac{1}{\kappa}$, where the LL mixing factor,  $ \kappa = (\frac{e^2}{l_B})/(\hbar \frac{e B}{m c})$, divides the expressions contrary to the case in the perturbation theory, and thus, again, we should be aware that we work with systems in which the LL mixing is considerable (no $ \kappa \rightarrow 0 $ limit).

But we can use the identified correspondence in ratios to conclude that the field theory correctly predicts the main features of a model interaction for Pfaffian. Namely,  it predicts negative values of three-body PPs, $V_M (\lambda) = \Lambda   \Delta_M =  (1/2 + \lambda ) \cdot 4/m  \cdot \Delta_M$,   for $M = 3, 5,$ and $6$ to be crucial for the establishment of Pfaffian according to the expression in Eq. (\ref{Int}), where we take  $\lambda  \lesssim - 1 $ in the Pfaffian case. This prediction is in a complete accordance with the numerical work in Ref. [\onlinecite{pp}] . In Fig. 3 of that work we see that the negative values of three-body PPs, with the specified ratios (based on the first order perturbation theory in the second LL), are crucial in obtaining Pfaffian state.  (We remind the reader of the comments in the paragraph below Eq. (\ref{gc}), that we model paired  states in a fixed, lowest LL and this model interaction is relevant for any fixed LL, including second LL.)  In the Table I we listed also calculated matrix element values for higher angular momenta  $(M = 7, 8, 9, 10 ) $ and (after the simple rescaling) they follow the basic trend of the first order perturbation theory for the second LL, which favors  Pfaffian and anti-Pfaffian states as analyzed in Ref. [\onlinecite{rez}].

The generation of three-body terms due to the LL mixing using the perturbation theory has a long history  [\onlinecite{mix1,mix2,mix3,mix4}]. It lead 
to the identification of relevant three-body parameters for an effective Pfaffian physics. Here we found a similar description of an effective interaction for Pfaffian using only field-theoretical arguments.

It is an interesting question whether the field theory may predict also a model interaction for anti-Pfaffian (or differentiate anti-Pfaffian from Pfaffian). The field theory can describe anti-Pfaffian as a Pfaffian pairing instability of Fermi liquid of composite holes [\onlinecite{bmf}], and for that case come up with the three-body interaction in (\ref{Int3explicit}). Threrefore we should consider that three-body interaction in a fixed LL, and apply PH transformation to get the model interaction for anti-Pfaffian.

Next, according to the formula (\ref{Int}) the field theory predicts that in the case of PH Pfaffian, $\lambda > 0$, positive three-body PPs are necessary for its establishment.
This is quite expected due to the role of negative ones in the establishment of Pfaffian and anti-Pfaffian (Fig. 3 in Ref. [\onlinecite{pp}], and Ref. [\onlinecite{rez}]), and, thus, the positive values will suppress the tendency to Pfaffian and anti-Pfaffian. 

Therefore,  the main feature of the model interaction for PH Pfaffian in a fixed LL (the projection of the PH Pfaffian) according to the field theory arguments is a series of positive three-body PPs, $V_M (\lambda)$ for ratios,  $\frac{V_{M}}{V_{M = 3}} = \frac{\Delta_{M }}{\Delta_{M = 3}}  $,
specified in the Table I.
Certainly the question remains whether a real system will slip into a Fermi liquid state and we will discuss this in the next section.

We may comment that according to the main features of the effective (model) interaction for PH Pfaffian (in a fixed LL) listed above, we do not expect more-than-three-body (additional) PPs to be relevant. The main features are derived on the basis of the BCS reduction we described in the Eq. (\ref{bHam}), i.e. the reduction we believe is a faithful description of paired states. On the other hand, on the basis of the perturbation theory in $\kappa$ we would expect also more-than-three-body PPs. If indeed the PH Pfaffian physics is present, for some $\kappa \gtrsim 1 $, it may be preceded  by a distinct phase for which the perturbation theory in $\kappa$ is valid.  On the other hand, the PH Pfaffian would be based on a non-perturbative (non-analytic) in $\kappa$ description of the LL mixing.

\section{Beyond the projection to a fixed Landau level}

We argued for main features of a model interaction for PH Pfaffian in a fixed LL. But we should also note that there are strong arguments that the projection will represent a gapless state:

(a) Let's assume that the projection (i.e. associated wave function) is PH symmetric . According to Refs. [\onlinecite{avm}] and [\onlinecite{mvm}] i.e. arguments in Section II of the Ref. [\onlinecite{avm}] based on Dirac composite fermion  (manifestly PH symmetric) theory (more precisely Bogoliubov description of the pairing of Dirac composite fermions which encapsulates the BCS ground state) such a state must be critical (gapless).

But we derived a model Hamiltonian (interaction) in a fixed LL that has an explicit three-body interaction which breaks PH symmetry and thus we should consider also the possibility that the projection is not PH symmetric.

(b) If the projection is not PH symmetric and represents a gapped state, the state based on the projection and its corresponding partner, a distinct phase, that we get by the PH exchange, have the same shift  i.e. an integer - a topological number that characterizes the state of the system on a curved background, such as a sphere. This is certainly not a sign for two distinct phases. Moreover, the numerical results in Ref. [\onlinecite{mish}] of the overlap of the projection and its partner under the PH exchange is so high for system sizes up to $N = 12$ despite the fact that the overlap must decay to 0 in the thermodynamic limit, irrespective of the presence of the PH symmetry.  (Thus either the projection is PH symmetric and we are back to the preceding case, or the state is gapless.)


Our model interaction for the PH Pfaffian in Eq. (\ref{Ham}) (or (\ref{Int})) is defined on the space organized by LLs, and a question is whether we will capture the nature and physics of the PH Pfaffian if we consider  only one LL for which the  rescaled magnitudes of the three-body PPs ( divided by  $\Lambda =  (1/2 + \lambda ) \cdot 4/m$),  are specified  in the Table I.  We can take
that the effective LL mixing parameter in this system is $ | 1/2 + \lambda |$.  Thus in  the case of Pfaffian when $ \lambda = - 1 $, we can stay in a fixed LL, while if $\lambda $ is of the same magnitude but opposite sign i.e. when we have the case of the PH Pfaffian with $ \lambda = 1 $, it seems we need to consider an additional LL.

To assess the role of higher LLs we concentrate on the three-body  interaction more precisely diagonal matrix elements of the three-body interaction when one, two or three electrons are in one higher LL  (the second LL).  We considered the wave functions that we get by applying  the raising operators $$a_i^\dag=\sqrt{2}\left(-\partial_{z_i}+z_i^* /4\right), $$ $ i = 1, 2, 3 $ to the 
lowest LL wave functions in Eq. (\ref{3wf}); we considered applying  (1)  $\frac{1}{\sqrt{3}}\left(a_1^\dag + a_2^\dag + a_3^\dag\right)$ (one electron of three electrons in the second LL - equivalent to a center of mass excitation), (2) 
$\frac{1}{3}\left(a_1^\dag a_2^\dag+a_2^\dag a_3^\dag+a_1^\dag a_3^\dag\right)$
(two electrons of three electrons in the second LL), and (3)  $a_1^\dag a_2^\dag a_3^\dag$ (all three electrons in the second LL). While calculating  these elements we encountered  ultraviolet divergences, because  of the limitations of the effective CS theory and its inability to capture short-range physics. Therefore we had to regularize the interaction in 
 Eq. (\ref{3V}) (of the effective CS description). Instead of $ |\mathbf{r}_3 - \mathbf{r}_1 |^{2}  \cdot  |\mathbf{r}_3 - \mathbf{r}_2 |^{2} $ in the denominator of  Eq. (\ref{3V}), we took
$  (|\mathbf{r}_3 - \mathbf{r}_1 |^{2} +  a^{2}) \cdot  (|\mathbf{r}_3 - \mathbf{r}_2 |^{2} +  a^{2}) $, where $a$ is a short-distance cut-off. We checked that if the denominator is modified into $ |\mathbf{r}_3 - \mathbf{r}_1 |^{2} \cdot  |\mathbf{r}_3 - \mathbf{r}_2 |^{2} + a^{4}, $ the values of implied PPs for $ a  \lesssim l_B $ do not change significantly.  The values of implied matrix elements, when $ a = 1 = l_B $ are given in the  tables below.

\begin{table}[h!]
\begin{center}
\caption{Matrix elements in the lowest Landau level with regularization.}
\begin{tabular}{|c||c|c|c|c|c|c|c|}
\hline
M & 3 & 5 & 6 & 7 & 8 & 9 & 10 \\
\hline\hline
$\Delta_M$ & $0.03309$ & $0.01944$ & $0.02487$ & $0.01186$ & $0.01592$ & 
\begin{tabular}{l|l}
$0.01946$ & \\
\hline
 & $0.00897$
\end{tabular} & $0.01047$ \\
\hline
$\frac{\Delta_M}{\Delta_{M=3}}$ & $1$ & $\sim 0.587$ & $\sim 0.752$ & $\sim 0.358$ & $\sim 0.481$ & 
\begin{tabular}{l|l}
$\sim 0.588$ & \\
\hline
 & $\sim 0.271$
\end{tabular} & $\sim 0.316$ \\
\hline
\end{tabular}
\end{center}
\end{table}

\begin{table}[h!]
\begin{center}
\caption{Matrix elements for states with two particles in the lowest Landau level and one particle in the second Landau level.}
\begin{tabular}{|c||c|c|c|c|c|c|c|}
\hline
M & 2 & 4 & 5 & 6 & 7 & 8 & 9 \\
\hline\hline
$\Delta_M$ & $0.03279$ & $0.01882$ & $0.02528$ & $0.01136$ & $0.01933$ & 
\begin{tabular}{l|l}
$0.01880$ & \\
\hline
 & $0.00844$
\end{tabular} & $0.01212$ \\
\hline
$\frac{\Delta_M}{\Delta_{M=2}}$ & $1$ & $\sim 0.574$ & $\sim 0.771$ & $\sim 0.346$ & $\sim 0.590$ & 
\begin{tabular}{l|l}
$\sim 0.573$ & \\
\hline
 & $\sim 0.257$
\end{tabular} & $\sim 0.370$ \\
\hline
\end{tabular}
\end{center}
\end{table}

\begin{table}[h!]
\begin{center}
\caption{Matrix elements for states  with one particle in the lowest Landau level and two particles in the second Landau level.}
\begin{tabular}{|c||c|c|c|c|c|c|c|}
\hline
M & 1 & 3 & 4 & 5 & 6 & 7 & 8 \\
\hline\hline
$\Delta_M$ & $0.02880$ & $0.02132$ & $0.02302$ & $0.01402$ & $0.01430$ & 
\begin{tabular}{l|l}
$0.01791$ & \\
\hline
 & $0.00880$
\end{tabular} & $0.01046$ \\
\hline
$\frac{\Delta_M}{\Delta_{M=1}}$ & $1$ & $\sim 0.740$ & $\sim 0.799$ & $\sim 0.487$ & $\sim 0.497$ & 
\begin{tabular}{l|l}
$\sim 0.622$ & \\
\hline
 & $\sim 0.305$
\end{tabular} & $\sim 0.363$ \\
\hline
\end{tabular}
\end{center}
\end{table}

\begin{table}[h!]
\begin{center}
\caption{Matrix elements in the second Landau level.}
\begin{tabular}{|c||c|c|c|c|c|c|c|}
\hline
M & 0 & 2 & 3 & 4 & 5 & 6 & 7 \\
\hline\hline
$\Delta_M$ & $0.02488$ & $0.02113$ & $0.01978$ & $0.01427$ & $0.01256$ & 
\begin{tabular}{l|l}
$0.01474$ & \\
\hline
 & $0.01012$
\end{tabular} & $0.01119$ \\
\hline
$\frac{\Delta_M}{\Delta_{M=0}}$ & $1$ & $\sim 0.849$ & $\sim 0.795$ & $\sim 0.573$ & $\sim 0.505$ & 
\begin{tabular}{l|l}
$\sim 0.592$ & \\
\hline
 & $\sim 0.407$
\end{tabular} & $\sim 0.450$ \\
\hline
\end{tabular}
\end{center}
\end{table}

When a value for certain matrix element is missing the numerical error was substantial.


\section{Discussion and conclusions}
We need to analyze more closely the role of parameter $ \lambda $. The sign of  $ \lambda $ in the composite fermion picture  (\ref{bHam}) determines the chirality of the underlying $p$-wave topological superconductivity of composite fermions, and  after the CS transformation into the electron picture,  in (\ref{Ham}), the parameter  $ \lambda $ determines whether we are in the Pfaffian, composite Fermi liquid (CFL)  [\onlinecite{hlr}], or PH Pfaffian phase. Even without the help of exact diagonalizations, we can come up with a schematic phase diagram of the electron system as a function of  $ \lambda$, see Fig. 1. Two insights lead to the phase diagram as a function of  $ \lambda $: (a) effective  three-body coupling is  $ (1/2 + \lambda) $  (this determines its sign), and (b) effective LL mixing parameter is  $| 1/2 + \lambda| $ (this determines how many LLs we need to include). (We consider two-body interactions, which are likely positive and monotonically decreasing with momenta, irrelevant for (PH) Pfaffian pairing. The paired states are expected to be stabilized by three-body interactions.)

\begin{center}
\begin{figure}[h!]
\includegraphics[scale=0.45]{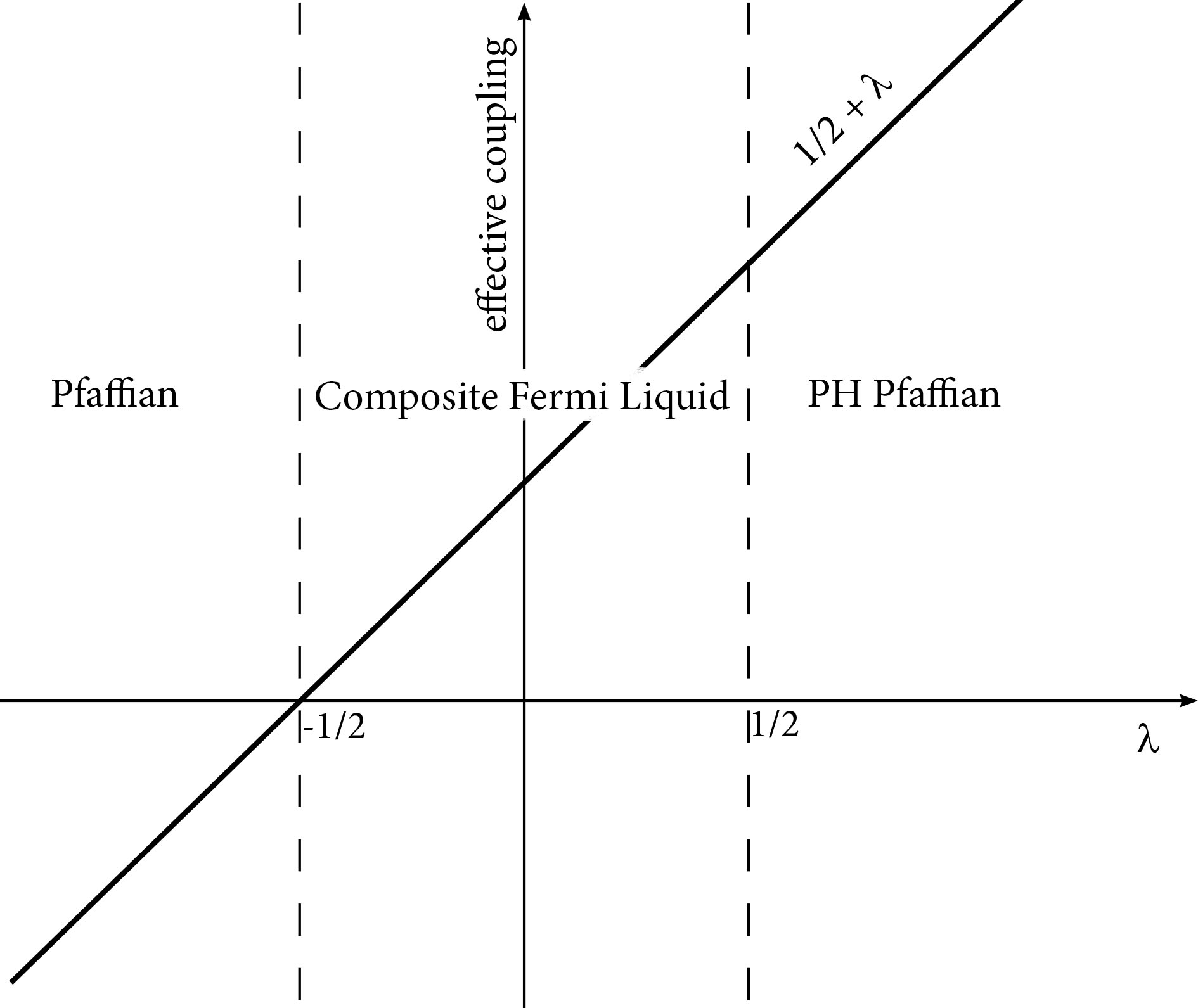}
\caption{Schematic phase diagram as the pairing parameter  $ \lambda $ is varied. }
\end{figure}
\end{center}

As we change parameter   $ \lambda$ from negative to positive values, more precisely for   $  \lambda   \gtrsim  - 1 $, we can estimate (due to the effective LL mixing  $\sim  | 1/2 + \lambda| $ ) that we need an extra LL to describe the electron system with underlying  BCS $p$-wave pairing of CFs at (around) $  \lambda = 1/2 $ . Prior to that value of $ \lambda $ , for 
$ - 1  \lesssim   \lambda < - 1/2 $, we expect a Pfaffian instability of electrons; this is based on our expectation (corroborated with numerics in   [\onlinecite{pp,rez}]) that the negative values of three-body interaction with specific ratios for lower angular momenta (Table I and Fig. 2)  will support the Pfaffian physics. The Pfaffian physics largely occurs in a fixed LL, and the ground state wave function of Pfaffian phase can be described by a completely holomorphic expression in a (fixed) lowest LL. The fixed LL physics is scale invariant in a special way; the characteristic length is only present in the exponential  factor which does not change as correlations in a fixed LL change. To describe the Pfaffian physics it is sufficient to stay in the lowest LL and use the (unregularized, negative values of ) PPs  described in Fig. 2. The physics does not depend on any length scale, and, though surprising at first sight that the field theory can come up with finite matrix elements, we can use them without any need to regularize. As we increase    $\lambda$, for $ - 1/2  <  \lambda  \lesssim  1/2 $, the effective three-body interaction is positive, because the value of the effective coupling, $(1/2 +  \lambda)$, changes sign to positive at $  \lambda = - 1/2 $. Thus in this effective description a phase transition  may occur at  $  \lambda = - 1/2 $. We expect an entrance into the (compressible) CFL phase.  For $ - 1/2  <  \lambda  \lesssim  1/2 $ the LL mixing,  $ \sim  | 1/2 + \lambda| $, is not large and we may consider also in this region only PPs of the lowest LL. At $ \lambda = 0$, in the composite fermion representation, as well electron representation, we have a Fermi liquid phase. Furthermore, for a whole interval,  $ - 1/2  <  \lambda  \lesssim  1/2 $, we expect a CFL phase (in the electron system),  because for $ \lambda = - 1/2$ there is an abrupt change in the sign of the three-body interaction accompanied with an oscillatory behavior in the positive values of PPs as a function of the total angular momentum of three fermions. The oscillatory behavior of the (positive) values of PPs might be a sign of the compressible correlations in the phase that we expect to be the CFL; a state of three fermions may reduce its angular momentum without a resistance (or significant increase in energy). Thus a (single) series  with positive oscillatory values of three-body PPs in the lowest LL (see Table 1 and Fig. 2)  may be a hallmark of the whole region,  $ - 1/2  <  \lambda  \lesssim  1/2 $, in which the topological pairing at weak coupling of composite fermions in (\ref{bHam}) is suppressed under the CS transformation into the electron representation. This system, at
 $ 0  < \lambda  \lesssim  1/2 $, as we will discuss  below, confined to a single LL, also represents a projection of the PH Pfaffian to a fixed LL - any attempt to confine the description of the PH Pfaffian in a fixed LL will produce a gapless state, state close to the CFL   [\onlinecite{avm,mish}].

\begin{center}
\begin{figure}[h!]
\includegraphics[scale=0.4]{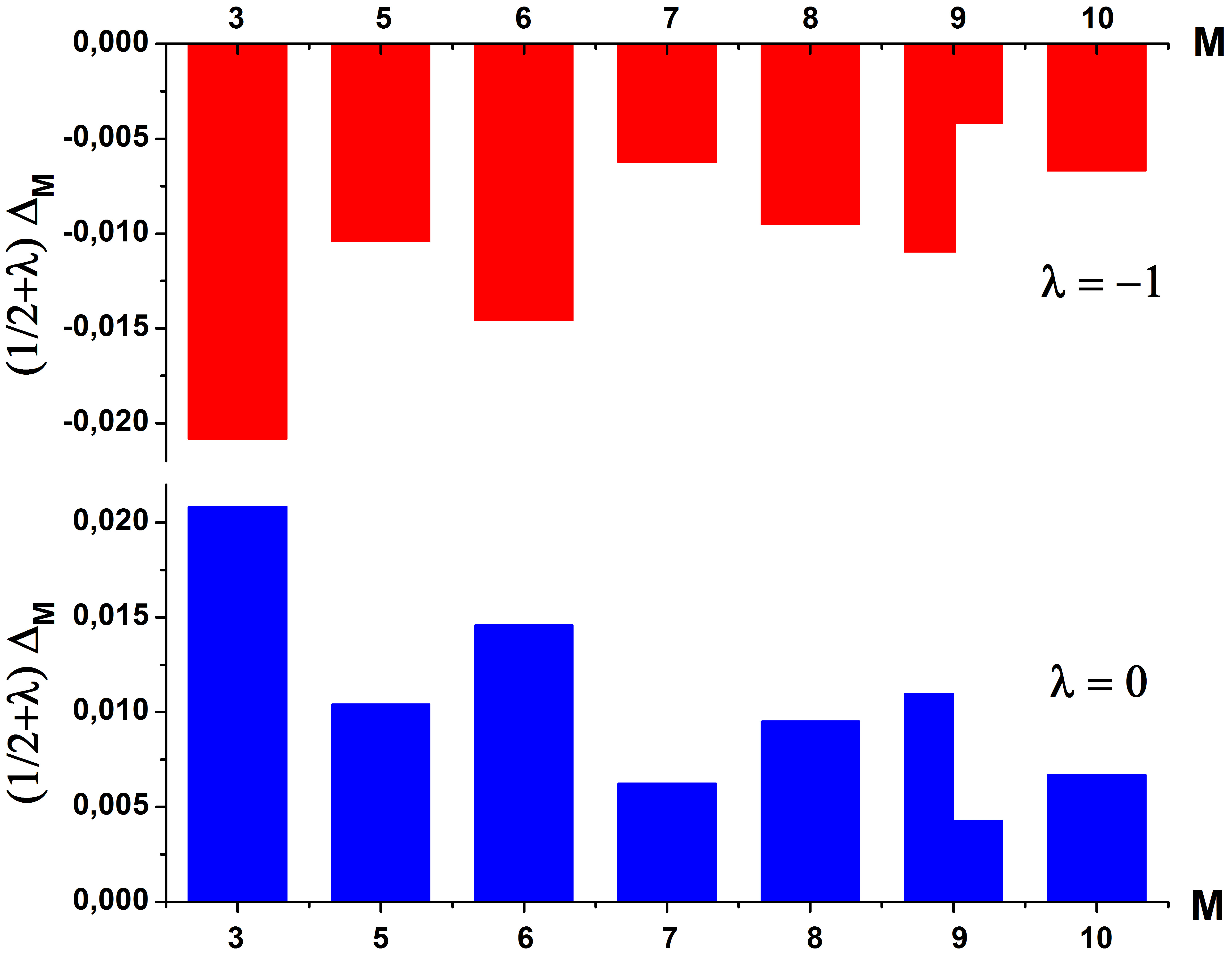}
\caption{Matrix elements of three body pseudo-potentials in the lowest Landau level for $\lambda=-1$ (above) and $\lambda=0$ (bottom).}
\end{figure}
\end{center}

But we should note and emphasize that the particular oscillatory behavior of PPs,  as described in the lower part of Fig. 2, with positive values, may promote the pairing necessary for PH Pfaffian, in which any three electrons can mostly correlate (efficiently minimize their energy) in the total angular momentum equal to $M = 7$, which is the characteristic angular momentum for PH Pfaffian pairing. (Recall that the characteristic angular momentum for Pfaffian is $ M = 5 $  [\onlinecite{src}].) Nevertheless, the compressible, Fermi-liquid-like behavior results from the projection to a fixed LL, due to phase-space constraints which prohibit the pairing. Namely, the expected leading term in the projection of the paired state is equal to zero, if the (unprojected) pairing function is  $ g(z) \sim  1/z^*$. We will come back to this point in the concluding remarks.

\begin{center}
\begin{figure}[h!]
\includegraphics[scale=0.4]{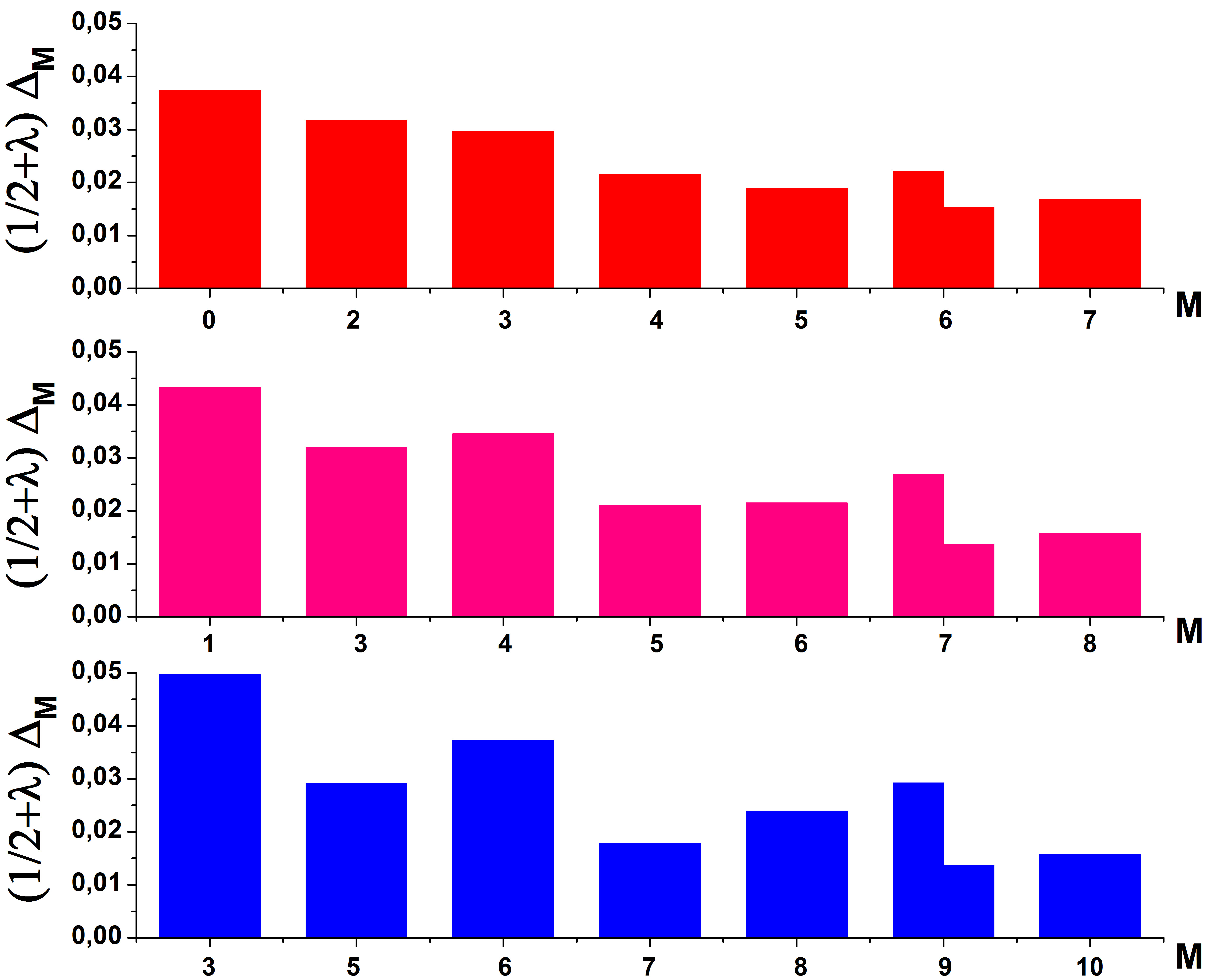}
\caption{Three body pseudo-potential matrix elements for $\lambda=1$  (PH Pfaffian case)  in the second Landau level (top), for states with two particles in the second Landau level and one in the lowest Landau level (middle), and in the lowest Landau level (bottom).}
\end{figure}
\end{center}

Because the LL mixing parameter in our model Hamiltonian,  (\ref{Ham}), can be estimated to be equal to  $|1/2 + \lambda| $, above  $  \lambda  \sim  1/2 $ we need to include at least one additional, extra LL (the second LL) to capture the underlying $p$-wave pairing state of the electron system. This brings the natural scale (for distance) - the magnetic length ($ l_B $)  as an external, fixed scale that we used to regularize the three-body PPs as described in the previous section. The regularized intra-LL PPs are given in that section. What we can notice is that if we confine our description to the lowest LL the PPs are still characterized by oscillatory behavior, see Table II and Fig. 3, and this can lead to the compressible state. On the other hand, interestingly, the three-body intra-LL PPs for the higher - second LL are characterized by monotonically decreasing (with total angular momentum) positive values -  Table V and Fig. 3. More importantly, in Table IV and Fig. 3, in the case of three electrons of which two are in the second LL, we see an abrupt decrease in the values of repulsive PPs, that occurs at $M = 5$ (effectively $M = 7$ in the lowest LL) - a characteristic angular momentum for the PH Pfaffian pairing ($ M = 5$ in the case of Pfaffian (in the lowest LL)). This  opens up a possibility for PH Pfaffian pairing correlations, which by definition are anti-holomorphic, to form and also exist in the higher band - second LL. At least two LLs are needed to establish the PH Pfaffian. This is not surprising given the fact that the Pfaffian antisymmetrized product of Cooper pairs with the projected pairing function $1/z^*$ to the LLL, $ g_{LLL} (z)  \sim  z $ is zero. The same pairing function, projected to the second LL is $ g_{sLL} (z, z^* )  \sim  ( | z |^2 - 4 ) \; z $, and thus the extra factor, $ ( | z |^2 - 4 )$ brings the (magnetic) length scale into description and enables a non-trivial pairing to develop and exist at short distances. But we should note that the values of calculated PPs for electronic correlations, do not lead immediatelly to an expectation for the existence of a gapped paired state; the transformation to the electron representation may lead to a compressible state with some pairing correlations due to the inclusion of the second LL. Further numerical investigations are necessary to probe the existence of a gapped state with the PH Pfaffian topological characterization, on the basis of the calculated PPs for the lowest two LLs. In this work we developed a general framework - a model interaction that can be used in the investigation of the PH Pfaffian. 

In the field-theoretical approach that we considered, besides using the mean-field (classical) equations of motions for fields in the effective composite fermion theory, we did not do any further approximations. Our approach clearly calls for a necessary inclusion of other LLs in numerical approaches, in the quest for PH Pfaffian. By working in a fixed LL, no matter how well the influence of other LLs is included, one can not access the PH Pfaffian pairing. We described the pertinent three-body parameters for two LLs.

\section{Acknowledgments}

We  would like to thank N. Regnault  and K. Pakrouski  for discussions.  This research
was supported by the Ministry of Education, Science, and
Technological Development of the Republic of Serbia under
Project ON171017, and by the Ministry of Science of Montenegro under Project SFS013454.

\appendix

\section{Statistical interaction}
In this Appendix the details of the calculations that lead from the statistical interaction in  Eq. (\ref{Vst})  to the Cooper channel expression in  Eq. (\ref{pairint})  will be explained.
If $\Psi_{cf} = \sum \frac{1}{\sqrt{V}} \exp\{i \mathbf{k} \cdot \mathbf{r}\} \; a_{\mathbf{k}}$ and
\begin{equation}
\mathbf{j}_{cf} (\mathbf{r})= \frac{1}{2 m} (\Psi_{cf}^\dagger (-i \mathbf{\nabla} \Psi_{cf}) + (i \mathbf{\nabla} \Psi_{cf}^\dagger) \Psi_{cf}),
\end{equation}
then
\begin{equation}
\mathbf{j}^{cf}_{\mathbf{p}}= \int d \mathbf{r} \exp\{i \mathbf{p} \cdot \mathbf{r}\} \;  \mathbf{j}_{cf} (\mathbf{r})= \frac{1}{2 m} \sum_{\mathbf{k}} a^\dagger_{\mathbf{p} +  \mathbf{k}} 
a_{\mathbf{k}} (2 \mathbf{k} + \mathbf{p}).
\end{equation}
Using the solutions in  (\ref{ax}) and  (\ref{ay}),  it follows that
\begin{equation}
\int \frac{\mathbf{j}_{cf} \cdot \mathbf{a}}{2} = \int d\mathbf{r} \int d\mathbf{r}^{'} \{  j_{x}^{cf}(\mathbf{r}) \frac{y - y^{'} }{|\mathbf{r} - \mathbf{r}^{'} |^2}  \rho(\mathbf{r}^{'}) -
 j_{y}^{cf}(\mathbf{r})  \frac{x - x^{'} }{|\mathbf{r} - \mathbf{r}^{'} |^2}  \rho(\mathbf{r}^{'} ) \}.
\end{equation}
If we introduce
\begin{equation}
\mathbf{j}_{cf}(\mathbf{r}) = \frac{1}{V} \sum_\mathbf{k} \exp\{i \mathbf{k} \mathbf{r}\} \mathbf{j}_{cf}(-\mathbf{k}),
\end{equation}
and 
\begin{equation}
\rho(\mathbf{r}) = \frac{1}{V} \sum_\mathbf{k} \exp\{- i \mathbf{k} \mathbf{r}\} \rho(\mathbf{k}),
\end{equation}
we can rewrite the above expression as
\begin{equation}
\int \frac{\mathbf{j}_{cf} \cdot \mathbf{a}}{2} =  \frac{1}{V} \sum_\mathbf{k} i \{ \frac{k_y}{|\mathbf{k}|^2} j^{cf}_x  (-\mathbf{k}) - \frac{k_x}{|\mathbf{k}|^2} j^{cf}_y  (-\mathbf{k}) \}  \rho(\mathbf{k}) (2 \pi ).
\end{equation}
With
\begin{equation}
\rho(\mathbf{k}) = \sum_{\mathbf{p}}  a^\dagger_{\mathbf{p} +  \mathbf{k}} 
a_{\mathbf{p}} ,
\end{equation}
the statistical interaction becomes
\begin{equation}
\int V_{st} = (-i) \frac{2 \pi }{m} \frac{1}{V} \sum_{\mathbf{k}, \mathbf{p}, \mathbf{q}} 2 \frac{ (q_x  k_y - q_y  k_x )}{|\mathbf{k}|^2}    a^\dagger_{\mathbf{p} +  \mathbf{k}} 
a_{\mathbf{p}}   a^\dagger_{-\mathbf{k} +  \mathbf{q}} 
a_{\mathbf{q}} .
\label{stint}
\end{equation}
If we consider only Cooper channel, i.e. $\mathbf{q} = - \mathbf{p}$,
\begin{equation}
V_{st}^{C}  = (-i) \frac{2 \pi }{m} \frac{1}{V} \sum_{\mathbf{k}, \mathbf{p}} (- 2) \frac{ (p_x  k_y - p_y  k_x )}{|\mathbf{k}|^2}    a^\dagger_{\mathbf{p} +  \mathbf{k}} 
a_{\mathbf{p}}   a^\dagger_{-\mathbf{k} -  \mathbf{p}} 
a_{-\mathbf{p}} .
\end{equation}
Introducing $ \mathbf{q} = \mathbf{k} +  \mathbf{p} $ we get Eq. (\ref{pairint}) in the main text.

To have a regularized, non-divergent statistical interaction, we should exclude the $ \mathbf{k} = 0 $ point in the summation, in (\ref{stint}), which amounts to an elimination of the part of  $ \mathbf{a}$ that describes
the external, constant magnetic field i.e.  $V_{st} =  - \mathbf{a}  \cdot \mathbf{j}_{cf} \rightarrow V_{st} ({\rm regularized}) =  - \delta \mathbf{a}  \cdot \mathbf{j}_{cf} $. This does not influence the Cooper channel description, and
we omitted this regularization in the Section II, but it is important for the discussion in the Section III.

\section{Two-body pseudo-potentials}

In this Appendix  the details of the calculation of the PPs for the two-body interaction,  $ \int d\mathbf{r} \;  \mathbf{a} \cdot \mathbf{j}_{el}$, in the lowest and second LL will be presented.
The projection of the first term in (\ref{Int}), 
\begin{equation}
 (1 + \lambda) \delta \mathbf{a} \mathbf{j}_{el} ,
\end{equation}
can be found by considering its second quantized form with field operators that belong (are projected) to a fixed LL. We consider
\begin{equation}
\delta \mathbf{a} \mathbf{j}_{el} = \frac{\phi_0}{\pi} \int d \mathbf{r}^{'} \frac{1}{2 i} \frac{\overline{j}_{el} (z - z^{'}) - j_{el} (\overline{z} - \overline{z}^{'})}{|z - z^{'}|^2} \delta \rho(\mathbf{r}^{'}),
\end{equation}
and thus
\begin{equation}
\int d \mathbf{r} \; \mathbf{a} \mathbf{j}_{el} = - \frac{1}{2 m}   \int d \mathbf{r}        \frac{\phi_0}{\pi} \int d \mathbf{r}^{'} \frac{\{
 [ \Psi^\dagger \frac{\partial}{\partial z} \Psi - \frac{\partial}{\partial z} \Psi^\dagger  \;  \Psi ]  (z - z^{'}) -
 [ \Psi^\dagger \frac{\partial}{\partial \overline{z}} \Psi - \frac{\partial}{\partial\overline{z}} \Psi^\dagger \;  \Psi ]                                                                                                                           (\overline{z} - \overline{z}^{'})
 \} }{|z - z^{'}|^2}  \Psi^\dagger  (\mathbf{r}^{'}) \Psi  (\mathbf{r}^{'}).
\end{equation}
By partial integration and using $ \frac{\partial}{\partial z} (\frac{1}{\overline{z}}) =  \frac{\partial}{\partial \overline{z}} (\frac{1}{z}) = 2 \pi \delta^2 (\mathbf{r} - \mathbf{r}^{'})$,
in the usual units,
\begin{equation}
\int d \mathbf{r} \; \mathbf{a} \mathbf{j}_{el} = - \frac{2}{m}   \int d \mathbf{r}   \int d \mathbf{r}^{'} \frac{\{
 [ \Psi^\dagger \frac{\partial}{\partial z} \Psi ]  (z - z^{'}) -
 [ \Psi^\dagger \frac{\partial}{\partial \overline{z}} \Psi  ]  (\overline{z} - \overline{z}^{'})
 \} }{|z - z^{'}|^2}  \Psi^\dagger  (\mathbf{r}^{'}) \Psi  (\mathbf{r}^{'}),
\end{equation}
with
\begin{equation}
\Psi(\mathbf{r}) = \sum  <\mathbf{r} | \Psi_n > \; a_n   =  \sum   { \cal M}_n   z^{n}  \exp\{- \frac{1}{4} |z|^2\} a_n   ,
\end{equation}
where
\begin{equation}
{ \cal M}_n  = {\cal N}_n  =  \frac{1}{\sqrt{2 \pi  2^n n!}},
\end{equation}
and $n = 0, 1, 2 , 3, \ldots $, in the lowest LL, and 
\begin{equation}
{ \cal M}_n  = \tilde{\cal N}_n   f_n ,
\end{equation}
where   $n = -1, 0, 1, 2 , 3, \ldots $,  and
\begin{equation}
\tilde{\cal N}_n  =  -  \frac{1}{\sqrt{2 \pi  2^{n+2} (n+1)!}},
\end{equation}
with
\begin{equation}
 f_n = (2 n + 2 -  \overline{z} z ) ,
\end{equation}
in the second LL.

In the lowest LL,
\begin{equation}
\int d \mathbf{r} \; \mathbf{a} \mathbf{j}_{el} =  \sum_{ \overline{n}, n,  \overline{m}, m}  - \frac{2}{m}   \frac{(2 \pi)^{2}}{2} 
 {\cal N}_ {\overline{n} }   \;  {\cal N}_n   \;   {\cal N}_ {\overline{m}}   \;  {\cal N}_m      \;    I(\overline{n}, n, \overline{m}, m)  \;
: a_ {\overline{n}}^{\dagger}  \;  a_n  \;  a_ {\overline{m}}^{\dagger}  \;   a_m  : ,
\end{equation}
and we have to use $\tilde{\cal N}_n $ 's instead of ${\cal N}_n $ 's, and calculate $ \tilde{I}(\overline{n}, n, \overline{m}, m)$ in the place of
$ I(\overline{n}, n, \overline{m}, m)$, for the effective interaction in the second LL.

We will consider the following diagonal elements with respect to states:  $  a_ {r}^{\dagger}  a_ {s}^{\dagger} | 0 > $ with $r = 0$ and $s = l$ in the lowest LL, and with
$r = -1$ and $s = l - 1$, in the second LL. In this way we can extract PPs, $ W_l^{2}$, $l = 1, 3, 5,  \ldots $, and  $ \tilde{W}_l^{2}$, $l = 1, 3, 5,  \ldots $, in the lowest and second LL, respectively :
\begin{equation}
  W_l^{2} =  - \frac{(2 \pi )^{2} }{m^{*}}   
 {\cal N}_ {l}^{2}   \;  {\cal N}_0^{2}   \;  \{  I(l, l, 0, 0)  + I(0, 0, l, l) - I(0, l, l, 0) - I(l, 0, 0, l) \},
\end{equation}
and 
\begin{equation}
  \tilde{W}_l^{2} =  - \frac{(2 \pi )^{2} }{m^{*}}   
 \tilde{\cal N}_ {l-1}^{2}   \;  \tilde{\cal N}_{-1}^{2}   \;  \{   \tilde{I}(l-1, l-1, -1, -1)  + \tilde{I}(-1, -1, l-1, l-1) - \tilde{I}(-1, l-1, l-1, -1) - \tilde{I}(l-1, -1, -1, l-1) \}.
\end{equation}
In the above formulae we used $m^{*} $ to denote the effective mass. The explicit expressions for $I$ and  $\tilde{I}$, with $  \Delta = m - \overline{m} - n +  \overline{n} $,  are
\begin{eqnarray}
 I(\overline{n}, n, \overline{m}, m) & =  & \int_0^{\infty} dr_1  r_1     \{  \int_0^{r_1} dr_2  r_2   \exp\{- \frac{1}{2} (|r_1 |^2 + |r_2 |^2 ) \}   \; 
r_2^{ \overline{m}  +  m +  \frac{\Delta}{2} }   r_1^{ \overline{n}  +  n -  \frac{\Delta}{2} - 2 }     (n - \frac{1}{4}  r_1^2 )    \nonumber   \\
&&    +    \int_{r_1}^{\infty} dr_2  r_2   \exp\{- \frac{1}{2} (|r_1 |^2 + |r_2 |^2 ) \}  \;    r_2^{ \overline{m}  +  m -  \frac{\Delta}{2}  - 2}   r_1^{ \overline{n}  +  n  + \frac{\Delta}{2}  }     ( - \frac{1}{4}  r_2^2 )     \},
\end{eqnarray}
and
\begin{eqnarray}
  \tilde{I}(\overline{n}, n, \overline{m}, m) & =  & \int_0^{\infty} dr_1  r_1     \{  \int_0^{r_1} dr_2  r_2   \exp\{- \frac{1}{2} (|r_1 |^2 + |r_2 |^2 ) \}  \;  
r_2^{ \overline{m}  +  m +  \frac{\Delta}{2} }   r_1^{ \overline{n}  +  n -  \frac{\Delta}{2} - 2 }    \times   \nonumber   \\
&&  (n   f_{\overline{n}} (r_1 )    f_n   (r_1 )       f_{\overline{m}}(r_2 )        f_m   (r_2 )       -     \frac{1}{4}  r_1^2    f_{\overline{n}} (r_1 )    \tilde{f}_n   (r_1 )       f_{\overline{m}}(r_2 )        f_m   (r_2 )         )    \nonumber   \\
&&    +  \int_{r_1}^{\infty} dr_2  r_2   \exp\{- \frac{1}{2} (|r_1 |^2 + |r_2 |^2 ) \}  \;    r_2^{ \overline{m}  +  m -  \frac{\Delta}{2}  - 2}   r_1^{ \overline{n}  +  n  + \frac{\Delta}{2}  }     f_{\overline{n}} (r_1 )    \tilde{f}_n   (r_1 )       f_{\overline{m}}(r_2 )        f_m   (r_2 )     ( - \frac{1}{4}  r_2^2 )     \},  \nonumber   \\
&&
\end{eqnarray}
where $ f_l (r) = 2 l + 2 - r^2  $, and $  \tilde{f}_l  (r )  = 2 l + 6 -  r^2 $.

Explicitly, for the lowest LL,
\begin{equation}
I_1 = I(l, l, 0, 0) = l!  \{-  \frac{1}{2} +  \frac{2^{l}}{2} -    \frac{2^{l}}{4}  \}   ,
\end{equation}
\begin{equation}
I_2 =  I(0, 0, l, l) = l!  \{ -    \frac{2^{l}}{4}  \} ,
\end{equation}
\begin{equation}
I_3 = I(0, l, l, 0) =  l!  \{-  \frac{1}{4} \},
\end{equation}
\begin{equation}
I_4 =  I(l, 0, 0, l) = l!  \{-  \frac{1}{4} \} .
\end{equation}
Therefore 
\begin{equation}
 W_l^{2}      \sim   ( - I_1 - I_2 + I_3 + I_4  )   = 0   ,
\end{equation}
for any $l = 1, 3, 5,  \ldots $

Explicitly, for the second LL,
\begin{equation}
 \tilde{I}_1 =  \tilde{I}(l-1, l-1, -1, -1) = l!  \{-  \frac{l^{2} + 2 l + 5 }{2} +  2^{l}   \}   ,
\end{equation}
\begin{equation}
 \tilde{I}_2 =   \tilde{I}(-1, -1, l-1, l-1) = l!  \{ -    \frac{l + 1}{2} + 2^{l}  \} ,
\end{equation}
\begin{equation}
 \tilde{I}_3 =  \tilde{I}(-1, l-1, l-1, -1) =  l!  \{-  \frac{3}{2} +  \frac{5}{4} l -  \frac{l^{2}}{4} +  2^{l+ 1} \},
\end{equation}
\begin{equation}
 \tilde{I}_4 =   \tilde{I}(l-1, -1, -1, l-1) = l!  \{-  \frac{l^{2} - 5 l + 6 }{4}    \}  .
\end{equation}
Therefore 
\begin{equation}
 \tilde{W}_l^{2}    =   \frac{ ( -  \tilde{I}_1  -  \tilde{I}_2  +  \tilde{I}_3 +  \tilde{I}_4  )  }{ 2^{l+ 3} l!}  = \frac{l}{ 2^{l+ 1}} ,
\end{equation}
for a given $l = 1, 3, 5,  \ldots $

\end{document}